\documentclass[twocolumn,aps,pra,amsmath,amssymb,showpacs]{revtex4}
\usepackage[latin9]{inputenc}
\usepackage{amsmath}
\usepackage{amssymb}
\usepackage{graphicx}
\usepackage{esint}

\makeatletter
\@ifundefined{textcolor}{}
{%
 \definecolor{BLACK}{gray}{0}
 \definecolor{WHITE}{gray}{1}
 \definecolor{RED}{rgb}{1,0,0}
 \definecolor{GREEN}{rgb}{0,1,0}
 \definecolor{BLUE}{rgb}{0,0,1}
 \definecolor{CYAN}{cmyk}{1,0,0,0}
 \definecolor{MAGENTA}{cmyk}{0,1,0,0}
 \definecolor{YELLOW}{cmyk}{0,0,1,0}
}

\usepackage{epsfig}\usepackage{amsfonts}\newcommand{\bra}[1]{\ensuremath{\left\langle {#1} \right|}}
\newcommand{\ket}[1]{\ensuremath{\left|  #1 \right\rangle}}
\newcommand{\uu}{\ket{\uparrow}_i\bra{\uparrow}_i}
\newcommand{\dd}{\ket{\downarrow}_i\bra{\downarrow}_i}
\newcommand{\eu}{\ket{e}_i\bra{\uparrow}_i}
\newcommand{\ed}{\ket{e}_i\bra{\downarrow}_i}
\newcommand{\ee}{\ket{e}_i\bra{e}_i}

\makeatother

\begin{document}

\title{Detuning Enhanced Cavity Spin Squeezing}

\author{Yan-Lei Zhang, $^{1,2}$ }

\author{Chang-Ling Zou, $^{1,2,3}$ }

\email{clzou321@ustc.edu.cn}

\author{Xu-Bo Zou, $^{1,2}$ }

\email{xbz@ustc.edu.cn}

\author{Liang Jiang, $^{3}$}

\author{Guang-Can Guo $^{1,2}$ }

\affiliation{$^{1}$ Key Laboratory of Quantum Information, University of Science
and Technology of China, Hefei, 230026, People's Republic of China; }

\affiliation{$^{2}$ Synergetic Innovation Center of Quantum Information \& Quantum
Physics, University of Science and Technology of China, Hefei, Anhui
230026, China}

\affiliation{$^{3}$ Department of Applied Physics, Yale University, New Haven,
CT 06511, USA}
\begin{abstract}
The unconditionally squeezing of the collective spin of an atomic
ensemble in a laser driven optical cavity {[}I. D. Leroux, M. H. Schleier-Smith,
and V. Vuleti\'{c}, Phys. Rev. Lett \textbf{104}, 073602 (2010){]}
is studied and analyzed theoretically. Surprisingly, we find that
the largely detuned driving laser can improve the scaling of cavity
squeezing from $S^{-2/5}$ to $S^{-2/3}$, where $S$ is the total
atomic spin. Moreover, we also demonstrate that the experimental imperfection
of photon scattering into free space can be efficiently suppressed
by detuning.
\end{abstract}


\maketitle
\emph{Introduction.-} Large ensembles of atoms are good platforms
for quantum information processing \cite{Hammerer2010,Braunstein2005,Korbicz2005},
as well as test beds of fundamental physics \cite{Wicht2002}, atomic
clocks \cite{Ludlow2008,NP14}, magnetometers \cite{Lukin,Wasilewski}
and gravitational wave detectors \cite{Mohr2008}. An important benchmark
for the protocols in high-precision measurements is spin squeezing
arising from the entanglement of atoms \cite{Kuzmich2000}. The squeezed
spin state (SSS) \cite{Kitagawa1993} is a quantum correlated state
with reduced fluctuations in one of the collective spin components,
which attracts considerable interest for both fundamental and practical
reasons.

Pumping the atomic ensemble by squeezed light \cite{Hald1999,Kuzmich1997,Fleischhauer2002,Fleischhauer2000}
has been proposed by quantum-state transfer from light to the atomic
spin. In this method, the degree of spin squeezing is determined by
the quality of the squeezed light, which is the source of spin squeezing.
The quantum nondemolition measurement \cite{Chaudhury2007,Fernholz2008,Takano2009,Inoue2013}
is another method of generating spin squeezing and has already been
performed by several groups \cite{Julsgaard2001}. The atomic ensembles
can be squeezed conditioned on the measurement results \cite{Leroux2012},
which are related to the performance of the detector. The last and
very promising system is the cavity squeezing \cite{Schleier2010,Leroux2010,Ueda1996,Kuzmich1998,Zhang2003,Takeuchi2005,Torre2013},
without requiring measurement of the light field. Cavity squeezing
relies on the off-resonant interaction between an ensemble of atoms
and a light field circulating in an optical resonator cavity, and
the ensemble spin imprints its quantum fluctuations on the light,
which acts back on the spin state to reduce those fluctuations. When
considering cavity squeezing \cite{Schleier2010,Leroux2010}, the
strong atom-cavity coupling is usually required and the effect of
detuning between the light and the cavity is not discussed. 
\begin{figure}
\includegraphics[width=0.9\columnwidth]{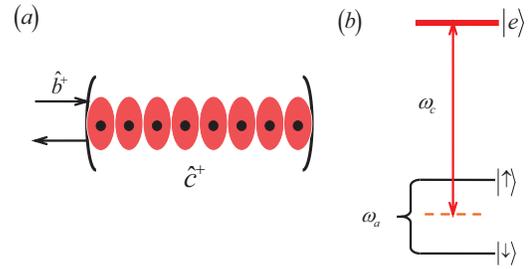} \protect\caption{(Color online) (a) Schematic illustration of an ensemble of atoms
uniformly coupled to optical cavity mode $c$, and a laser field is
driving the cavity. (b) Energy diagram of atom, the transitions between
lower states ($\left|\downarrow\right\rangle $ and $\left|\uparrow\right\rangle $)
and excited state $\left|e\right\rangle $ are coupled to cavity mode.}

\label{fig:setup} 
\end{figure}

In this paper, we theoretically study the detuning dependence of cavity
spin squeezing for the experimental scheme demonstrated in Ref.\cite{Leroux2010}
(Fig. 1a). Comparing with the near resonance case \cite{Schleier2010},
it is surprising to find that the scaling of cavity squeezing on atom
number can be significantly improved from $S^{-2/5}$ to $S^{-2/3}$
for large detuning. In addition, we find that the spin squeezing will
be enhanced if the atoms are weakly coupled to the cavity or the laser
detuning is very large. From our numerical solutions and analytical
analysis, the large detuning is very important as the squeezing originates
from the laser induced spin state dependent geometry phase \cite{Puri2012,Cirac}.
Finally, we study the influence of scattering of photon into free
space due to imperfect Raman scattering, and demonstrate that the
optimal spin squeezing can be obtained with appropriate detuning.
This improvement of spin squeezing by detuning is very feasible for
experiments, without the requirement of preparation or post-selection
of photon state. The detuning enhanced cavity spin squeezing can also
be applied to other systems, such as nitrogen-vacancy centers in diamond,
to prepare SSS for quantum metrology.

\emph{Model.-} The system (Fig. 1b) is an ensemble of $N$ identical
three-level atoms trapped inside an optical Febry-Pérot cavity. There
are two stable ground states $\left|\uparrow\right\rangle $ and $\left|\downarrow\right\rangle $,
which are coupled to the excited state $\left|e\right\rangle $ via
optical transitions of frequencies $\omega_{c}\pm\omega_{a}/2$. The
cavity resonance frequency $\omega_{c}$ is chosen so that the detunings
to transitions $\left|\uparrow\right\rangle \leftrightarrow\left|e\right\rangle $
and $\left|\downarrow\right\rangle \leftrightarrow\left|e\right\rangle $
are opposite in sign but having the same magnitude $\Delta=\omega_{a}/2$.
For simplicity, we only consider the case where the two transitions
have equal single-photon Rabi frequency $2g$ and all atoms are uniformly
coupled to the cavity. The Hamiltonian of the system reads ($\hbar=1$)
\begin{align}
H_{\mathrm{cav}}= & \omega_{c}c^{\dagger}c+\sum_{i=1}^{N}\Big{(}\frac{\omega_{a}}{2}\left[\uu-\dd\right]+\omega_{c}\ee\nonumber \\
 & +g\left[c\eu+c\ed+H.c.\right]\Big{)}.
\end{align}
Here, $c$ and $c^{\dagger}$ are the photon annihilation and creation
operators for the cavity mode, and the index $i$ labels the individual
atoms. As we are interested in the linear and dispersive regime of
atom-field interactions, we assume the excited state population is
negligible. The assumption requires a large detuning $|\Delta|\gg\kappa,\Gamma,g$
and sufficiently low intracavity photon number $\left\langle c^{\dagger}c\right\rangle \ll(\Delta/g)^{2}$,
where $\kappa$ is the cavity linewidth, $\Gamma$ is the excited
state decay rate. After adiabatically eliminating the excited state
of atom and considering external continuum fields \cite{Collett1984,Gardiner1985},
we obtain the effective Hamiltonian for the system
\begin{align}
H_{\mathrm{eff}}= & (\delta+\Omega S_{z})c^{\dagger}c+\int_{-\infty}^{\infty}\omega b_{\omega}^{\dagger}b_{\omega}d\omega\nonumber \\
 & +\sqrt{\kappa}\left[\beta_{\mathrm{in}}^{\ast}(t)c+c^{\dagger}\beta_{\mathrm{in}}(t)\right]\nonumber \\
 & +\sqrt{\frac{\kappa}{2\pi}}\int_{-\infty}^{\infty}\left(b_{\omega}^{\dagger}c+c^{\dagger}b_{\omega}\right)d\omega,
\end{align}
where $\delta=\omega_{c}-\omega_{l}$ is the detuning between the
resonator mode and the driving light, $S_{z}=\frac{1}{2}\sum_{i=1}^{N}(\uu-\dd)$
is the $z$-component of the total spin, $\Omega=2g^{2}/|\Delta|$
is the dispersive frequency shift due to spin-photon interaction,
$\beta_{\mathrm{in}}(t)$ is the driving, and $b_{\omega}\ (b_{\omega}^{\dagger})$
is the annihilation $($creation$)$ operator of the continuum.

Under coherent laser driving, the intracavity field is the coherent
state with spin-dependent phase shift. Assume the system is in state
\begin{equation}
\left|\psi\right\rangle =\sum_{m}C_{m}\left|\varphi_{m}(t),m\right\rangle \prod_{\omega}\left|\beta_{\omega,m}(t)\right\rangle ,
\end{equation}
Where $m$ is the quantum number associated with $S_{z}$, which is
conserved during the evolution, $\left|\varphi_{m}(t)\right\rangle $
is the cavity photon state and the $\left|\beta_{\omega,m}(t)\right\rangle $
is the state of continuum. By solving the Schrodinger equation $i\frac{\partial}{\partial t}\left|\psi\right\rangle =H_{\mathrm{eff}}\left|\psi\right\rangle $,
the time dependent intracavity field \cite{Puri2012} is 
\begin{equation}
\varphi_{m}(t)=-i\sqrt{\kappa}\int_{0}^{t}\beta_{\mathrm{in}}(t^{'})e^{-i(\delta+\Omega m)(t-t^{'})}e^{-\kappa(t-t^{'})/2}dt^{'},
\end{equation}
and the continuum modes are 
\begin{align}
\left|\beta_{\omega,m}(t)\right\rangle  & =e^{-i\int_{0}^{t}\mathrm{Re}(\sqrt{\kappa}\beta_{\mathrm{in}}^{\ast}(t')\varphi_{m}(t'))dt'}\left|\beta'_{\omega,m}(t)\right\rangle ,\\
\beta'_{\omega,m}(t) & =-i\sqrt{\frac{\kappa}{2\pi}}\int_{0}^{t}\varphi_{m}(t')e^{-i\omega(t-t')}dt'.
\end{align}

In general, the cavity photon, continuum and atomic spin states are
entangled {[}Eq. (3){]}. If the output field is not measured, the
density matrix of cavity photon and the atomic spin can be written
as 
\begin{equation}
\rho_{\mathrm{in,atom}}=\sum_{m,n}C_{m}C_{n}^{\ast}e^{\phi_{m,n}(t)}\left|\varphi_{m}(t),m\right\rangle \left\langle \varphi_{n}(t),n\right|
\end{equation}
by tracing the continuum modes out, where 
\begin{align}
\phi_{m,n}(t)= & -i\int_{0}^{t}\sqrt{\kappa}\mathrm{Re}(\beta_{\mathrm{in}}^{\ast}(t')\varphi_{m}(t')-\beta_{\mathrm{in}}(t')\varphi_{n}^{\ast}(t'))dt'\nonumber \\
 & -\kappa\int_{0}^{t}|\varphi_{n}(t')|^{2}dt'/2-\kappa\int_{0}^{t}|\varphi_{m}(t')|^{2}dt'/2\nonumber \\
 & +\kappa\int_{0}^{t}\varphi_{n}^{\ast}(t')\varphi_{m}(t')dt'.
\end{align}

The spin squeezing is evaluated by squeezing parameter \cite{Kitagawa1993}
\begin{equation}
\xi_{s}^{2}=\frac{min\left(\Delta S_{\vec{n}_{\perp}}^{2}\right)}{S/2},
\end{equation}
Where $\Delta S_{\vec{n}_{\perp}}^{2}$ is the variance of spin operators
along direction perpendicular to the mean-spin direction $\vec{n_{0}}=\frac{\vec{s}}{|\langle\vec{s}\rangle|}$,
which is determined by the expectation values $\left\langle S_{\alpha}\right\rangle $,
with $\alpha\in\left\{ x,y,z\right\} $. For an atomic system initialized
in a coherent spin state (CSS) \cite{Arecchi1972} along the $x$
axis, satisfying $S_{x}\left|\psi\left(0\right)\right\rangle _{\mathrm{atom}}=S\left|\psi\left(0\right)\right\rangle _{\mathrm{atom}}$,
we have $C_{m}=2^{-S}\sqrt{\frac{(2S)!}{(S-m)!(S+m)!}}$ and $\Delta S_{\vec{n}_{\perp}}^{2}=S/2$.
Thus, for squeezed spin states we have $\xi_{s}^{2}<1$.

\begin{figure}[htbp]
\center \includegraphics[width=0.9\columnwidth]{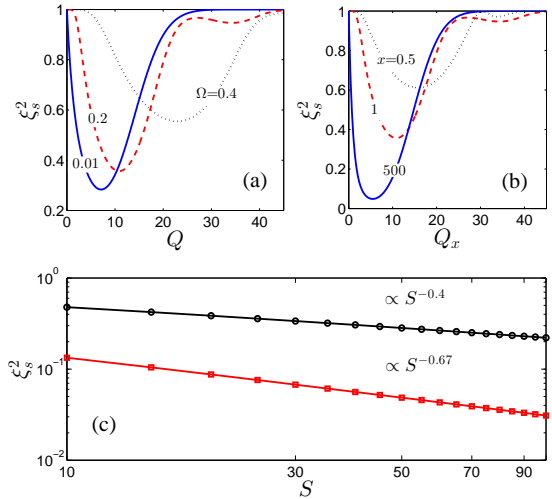} \protect\caption{(Color online). (a) The squeezing parameter $\xi_{s}^{2}$ as a function
of shearing strength $Q$ for $S=50$, $\delta=-\kappa/2$ and various
coupling $\Omega=0.01,\ 0.2,\ 0.4\ \mathrm{MHz}$. (b) The squeezing
parameter $\xi_{s}^{2}$ as a function of shearing strength $Q_{x}$
for $S=50$, $\Omega=0.2\ \mathrm{MHz}$ and various detuning $\delta=-x\kappa/2$,
$x=0.5,\ 1,\ 500$. (c) The optimal squeezing parameter $\xi_{s}^{2}$
as a function of the atomic spin $S$ for $\Omega=0.01\ \mathrm{MHz}$
and the detuning $\delta=-x\kappa/2$, $x=1$ (black), $x=\ 500$
(red). The other parameters are $\kappa=4\ \mathrm{MHz}$ and $\beta_{0}=1$. }
\end{figure}

\emph{Detuning enhanced squeezing.-} Now, we study the cavity spin
squeezing with continuous drive $\beta_{in}(t)=i\sqrt{\kappa}\beta_{0}$
with a small detuning $\delta=-\kappa/2$. For easier illustration,
it is useful to introduce the dimensionless shearing strength \cite{Schleier2010}
\begin{equation}
Q=\frac{4S|\beta_{0}|^{2}\Omega^{2}t}{\kappa},
\end{equation}
which is proportional to the transformation degree from the optical
field to the atomic spin. In Fig. 2 (a), we plot the spin squeezing
parameter $\xi_{s}^{2}$ as a function of shearing strength $Q$ for
various coupling $\Omega$. It clearly shows that the spin squeezing
parameter has a minimal value for certain optimal $Q$, and it takes
longer time for smaller coupling $\Omega$. The minimal value of spin
squeezing parameter increases with the coupling $\Omega$, because
there are higher order effects associated with $\Omega$ that will
limit the squeezing.

To study the effect of the detuning $\delta$ on spin squeezing, we
set $\delta=-x\kappa/2$, and the dimensionless shearing strength
can be generalized as

\begin{equation}
Q_{x}=4Qx/(1+x^{2})^{2},
\end{equation}
 In Fig. 2(b), we plot the spin squeezing parameter $\xi_{s}^{2}$
as a function of shearing strength $Q_{x}$ for various detuning $\delta$
with fixed coupling strength $\Omega=0.2\ \mathrm{MHz}$. The spin
squeezing can be enhanced for both red and blue large detuning $\delta$.
Since the larger detuning means that the driving light is hard to
enter into the cavity, the larger input power or longer interaction
time is required. It can be seen from Figs. 2 (a) and (b) that the
atomic spin can squeezed more than once until the atomic spin is fully
uncorrelated. This oscillation behavior is due to the competition
between the effective spin squeezing interaction, higher order effects
and decoherence.

In Fig. 2(c), we plot the optimal spin squeezing as a function of
the number of spins $S$, and the optimal spin squeezing is the minimum
value of $\xi_{s}^{2}(Q_{x})$. The black line shows the optimal squeezing
parameter $\xi_{s}^{2}\propto S^{-2/5}$ with the small detuning $\delta=-\kappa/2$,
as obtained in Ref. \cite{Schleier2010}. When we chose the large
detuning $\delta=-250\kappa$, the optimal squeezing parameter is
obtained as the red line, which satisfies $\xi_{s}^{2}\propto S^{-2/3}$.
Obviously, the spin squeezing is greatly enhanced by the detuning,
approaching the fundamental limitation of the one-axis spin squeezing
\cite{Kitagawa1993}. 

\emph{Mechanism.- }The Hamiltonian Eq. (2) implies that the atom-photon
interaction induces a spin state dependent geometric phase $\int dt\langle\varphi_{m}(t),m|\frac{\partial}{\partial t}\left|\varphi_{m}(t),m\right\rangle $
\cite{Puri2012,Cirac}. The spin squeezing is caused by the accumulated
geometric phase difference $\phi_{m,n}$ between the different spin
states $|m\rangle$ and $|n\rangle$. For continuous laser driving
and long interaction time $t\gg\kappa^{-1}$, the intracavity field
transient behavior can be neglected. The steady cavity field for detuning
$\delta=-x\kappa/2$ can be written as 
\begin{equation}
\varphi_{m}=\frac{\kappa\beta_{0}}{\kappa/2+i(\delta+\Omega m)}.
\end{equation}
From Eq. (8), we solve the phase factor as 
\begin{align}
\phi_{m,n}(t)=i & \frac{|\varphi_{m}|^{2}|\varphi_{n}|^{2}\Omega^{2}t}{\kappa\beta_{0}^{2}}\times\left\{ \frac{\frac{\kappa^{2}}{4}+\delta^{2}}{\Omega\kappa}\left(n-m\right)\right.\nonumber \\
 & \left.+\frac{\delta}{\kappa}\left(n^{2}-m^{2}\right)+\frac{\Omega}{\kappa}nm\left(n-m\right)+i\frac{\left(n-m\right)^{2}}{2}\right\} .
\end{align}
The first term accounts for the coefficient that approximately proportional
to $Q_{x}$, and the terms within the brace are the linear, quadratic
and higher order couplings of spin $z$-component. The quadratic term
corresponding to spin squeezing interaction $S_{z}^{2}$, while the
last two terms give rise to disorder and decoherence of spin states.
It's obvious that the detuning is essential in the cavity induced
spin squeezing, as there is no squeezing at all for zero detuning
$\delta=0$. The parameters $\frac{\delta}{\kappa}$ should be as
large as possible to make the squeezing effect outperform the undesired
effects, i.e. $\frac{\delta}{\kappa}\gg1$ and $\frac{\delta}{\kappa}\gg\frac{\Omega}{\kappa}$
should be satisfied. This can explain the results  the dependence
of optimal spin squeezing on $\delta$ and $\Omega$ shown in Figs.
2(a) and 2(b): (1) For very large $Q$ or $Q_{x}$, the disorder and
decoherence dominate over the coherent process. (2) Larger $\delta$
helps to suppress both disorder and dissipation. (3) Smaller $\Omega$
can suppress the high order terms, thus can enhance the squeezing.

For more intuitive understanding, we obtain the spin squeezing parameter
$\xi_{s}^{2}$ from the Heisenberg equation \cite{Schleier2010} under
certain approximation 
\begin{align}
\frac{\Omega}{\kappa}|Sz|\frac{1+|x|}{1+x^{2}} & \leq\frac{\Omega}{\kappa}\sqrt{S/2}\frac{1+|x|}{1+x^{2}}\ll1,\\
1\ll & |Q_{x}|\ll S,\\
\xi_{s}^{2}=\frac{1}{Q_{x}^{2}}+ & \frac{2}{Q_{x}x}+\frac{Q_{x}^{4}}{24S^{2}},\ x\neq0.
\end{align}
When $(5/2)^{5/4}12^{-1/4}S^{-1/2}\leq x\ll12^{1/6}S^{1/3}$, we obtain
the optimal cavity squeezing $\xi_{s,min}^{2}=(5/2)12^{-1/5}S^{-2/5}x^{-4/5}$
at the point $Q_{x}=12^{1/5}S^{2/5}x^{-1/5}$. When the detuning is
very large $x\gg12^{1/6}S^{1/3}$, the squeezing limit is $\xi_{s,min}^{2}=(3/2)12^{-1/3}S^{-2/3}$
with $Q_{x}=12^{1/6}S^{1/3}$. The detuning is the source of the effect
nonlinear interactions between the atomic spin and the optical mode,
and the large detuning means that the $1/Q_{x}^{2}$ is the main factor
of spin squeezing rather than the part $2/(Q_{x}x)$. We can improve
the scaling of cavity squeezing to $(3/2)12^{-1/3}S^{-2/3}$ with
sufficient detuning. 
\begin{figure}[htbp]
\center \includegraphics[width=0.9\columnwidth]{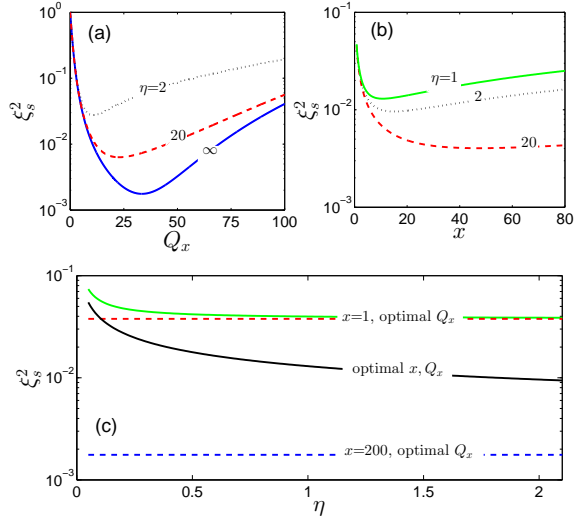} \protect\caption{(Color online). (a) The squeezing parameter $\xi_{s}^{2}$ as a function
of $Q_{x}$ for fixed detuning $x=200$, and single-atom cooperativity
$\eta$=$2$, $20$, $\infty$. (b) The optimal squeezing parameter
$\xi_{s}^{2}$ as a function of detuning $x$ for the various cooperativity
$\eta=1,\ 2,\ 20$. (c) The solid lines are optimal squeezing parameter
$\xi_{s}^{2}$ as a function of the cooperativity $\eta$ for the
fixed detuning $x=1$ (green) and optimized detuning (black), and
the dashed lines are results for ideal condition $\eta=\infty$ for
$x=1$ (red) and $x=200$ (blue). The atomic spin is $S=10^{4}$.}
\end{figure}

\emph{Imperfections.-} In previous studies, we have neglected the
scattering of photon into free space, which is an unavoidable process
that deteriorates squeezing performance \cite{Leroux2012}. Any atoms
scattering photon into free space will acquire a random phase, so
that it no longer contributes to the mean spin length. The Raman transitions
$|\uparrow\rangle\rightarrow|e\rangle\rightarrow|\downarrow\rangle$
or $|\downarrow\rangle\rightarrow|e\rangle\rightarrow|\uparrow\rangle$
reduce the correlation between the time average $\overline{S_{z}}$
during the cavity squeezing process. The average photon number emitted
into free space per atom is given by 
\begin{equation}
R_{x}=Q_{x}(1+x^{2})/(8xS\eta),
\end{equation}
which depends on the single-atom cooperativity $\eta=4g^{2}/(\kappa\Gamma)$.
This expression indicates that very large collective cooperativity
$S\eta\gg1$ is required to suppress the scattering of cavity photon
into free space. We extend the solution previously obtained in \cite{Leroux2012}
to the large detuning, and obtain the spin squeezing parameter: 
\begin{equation}
\xi_{s}^{2}=\frac{\left\langle \widetilde{S_{y}}^{2}\right\rangle +\left\langle S_{z}^{2}\right\rangle -\sqrt{\left\langle \widetilde{S_{y}}^{2}\right\rangle -\left\langle S_{z}^{2}\right\rangle +W^{2}}}{S},
\end{equation}
where $W=\left\langle \widetilde{S_{y}}S_{z}+S_{z}\widetilde{S_{y}}\right\rangle $
and the mean value of spin operators $\widetilde{S}$ are solved approximately
in the rotating frame as 
\begin{align}
 & \left\langle \widetilde{S_{y}}\right\rangle =\left\langle S_{z}\right\rangle =0,\left\langle S_{z}^{2}\right\rangle =\frac{S}{2},\\
 & \left\langle \widetilde{S_{y}}^{2}\right\rangle =\frac{S}{2}\left[1+Se^{-4R_{x}}\left(1-e^{-U}\right)\right],\\
 & \left\langle \widetilde{S_{y}}S_{z}+S_{z}\widetilde{S_{y}}\right\rangle =S\left(1-R_{x}\right)Q_{x}e^{-V},
\end{align}
with parameters $U=\frac{2Q_{x}}{xS}+\frac{Q_{x}^{2}\left(1-2R_{x}/3\right)}{S}$,
$V=\frac{Q_{x}}{2xS}+2R_{x}+\frac{Q_{x}^{2}\left(1-2R_{x}/3\right)}{4S}$. 

Although $\xi_{s}^{2}$ is a complicated function of $\eta$, $x$
and $Q_{x}$ due to imperfection, the spin squeezing can be optimized
for a given $\eta$ by adjusting the laser detuning and pump power
and interacting time. Fig. 3(a) shows the squeezing parameter $\xi_{s}^{2}$
as a function of $Q_{x}$ for various values of the cooperativity
$\eta$ and fixed large detuning $x=200$. And in Fig. 3(b), the optimized
spin squeezing parameters for certain $Q_{x}$ is calculated against
detuning $x$ for given cooperativity $\eta$. These results indicate
that the squeezing parameter is very sensitive to the value of the
cooperativity $\eta$, and better spin squeezing can be achieved for
larger $\eta$ and appropriate detuning $x$. Shown in the Fig. 3(c)
is the optimal squeezing parameter $\xi_{s}^{2}$ as a function of
the $\eta$. Green and black solid lines are the results for fixed
detuning ($x=1$) and optimized detuning. With increasing $\eta$,
$\xi_{s}^{2}$ is reduced and trend to be saturated at certain value.
Compared with the fixed detuning, the optimal detuning is always better,
indicating that the detuning regulation can efficiently enhance the
spin squeezing. When the cooperativity is not too small $\eta>0.1$,
the squeezing by optimized detuning can be even better than the result
of fixed detuning with $\eta=\infty$.

To lowest order expansion of $R_{x}\ll1$ and ignoring curvature effects
for the moment, the asymptotic solution of the squeezing parameter
{[}Eq. (18){]} can be written as 
\begin{equation}
\xi_{s}^{2}=Q_{x}^{-2}+\frac{2}{Q_{x}x}+\frac{Q_{x}(x^{2}+1)}{6xS\eta}.
\end{equation}
When the $\delta$ is very small, the squeezing variance suppressed
by the square of the shearing strength is neglected. Consequently,
there exist an optimum shearing strength $Q_{scatt}=\sqrt{12S\eta/(x^{2}+1)}$,
to achieve the optimum squeezing $\xi_{s}^{2}=\sqrt{\frac{4(x^{2}+1)}{3S\eta x^{2}}}$.
For very large detuning that satisfies $x\gg12^{1/6}S^{1/3}$, we
have optimum squeezing $\xi_{s}^{2}=3\left(\frac{1+x^{2}}{12xS\eta}\right)^{2/3}$
for the shearing strength $Q_{scatt}=\left(\frac{12xS\eta}{1+x^{2}}\right)^{1/3}$.
The squeezing is thus possible even for very weakly coupled resonator
and atoms with single photon-atom coupling cooperativity $\eta\ll1$,
as long as the collective cooperativity $S\eta\gg1$. 

\emph{Conclusion.-} We have theoretically analyzed the experimental
method to squeeze the collective spin of an atomic ensemble in a driven
optical cavity unconditionally. We find that strong atom-cavity coupling
weakens the spin squeezing and the large detuned laser driving can
improve the scaling of spin squeezing to $S^{-2/3}$ , which is the
ultimate limit of the ideal one-axis twisting spin squeezing. The
imperfection of light scattering into free space can be efficiently
suppressed by optimal detuning, which can be tested experimentally
and may further improve the sensitivity of quantum metrology based
on the SSS.

{\em Acknowledgments.} We thank M. H. Schleier-Smith for useful
discusions. This work was funded by National Natural Science Foundation
of China (Grant No. 11074244 and Grant No. 11274295), 973project (2011cba00200).
LJ acknowledges support from the Alfred P Sloan Foundation, the Packard
Foundation, the AFOSR-MURI, the ARO, and the DARPA Quiness program.

\end{document}